\documentclass[aps,pra,twocolumn,superscriptaddress,amssymb,amsmath,showpacs]{revtex4}

\usepackage{graphicx}

\pdfoutput=1

\begin{document}

\preprint{LA-UR-07-7306}

\title{Laserless trapped-ion quantum simulations without spontaneous scattering using microtrap arrays}


\author{J. Chiaverini}
\email[e-mail:\ ]{johnc@lanl.gov}
\affiliation{Applied Modern Physics Group, MS D454, Los Alamos National Laboratory, Los Alamos, New Mexico 87545, USA}

\author{W. E. Lybarger, Jr.}
\affiliation{Applied Modern Physics Group, MS D454, Los Alamos National Laboratory, Los Alamos, New Mexico 87545, USA}
\affiliation{Department of Physics and Astronomy, University of California, Los Angeles, California 90095, USA}


\date{\today}

\begin{abstract}

We propose an architecture and methodology for large-scale quantum simulations using hyperfine states of trapped ions in an arbitrary-layout microtrap array with laserless interactions.  An ion is trapped at each site, and the electrode structure provides for the application of single and pairwise evolution operators using only locally created microwave and radio-frequency fields.  The avoidance of short-lived atomic levels during evolution effectively eliminates errors due to spontaneous scattering; this may allow scaling of quantum simulators based on trapped ions to much larger systems than currently estimated. Such a configuration may also be particularly appropriate for one-way quantum computing with trapped-ion cluster states.

\end{abstract}


\maketitle


\section{Introduction\label{sec_intro}}

In his early proposal for quantum computing (QC), Feynman noted that the simulation of an interacting collection of quantum subsystems is complicated by the exponential scaling of the problem with system size~\cite{feynman82a}.  This is the basis for the technological fact that the current record for the number of interacting quantum spin-1/2 systems (or qubits) simulatable without approximation on a classical supercomputer is 36~\cite{deraedt07a}.  However, mapping the Hamiltonian of interest onto a more easily manipulated quantum system avoids this exponential scaling~\cite{lloyd96a}.  If a faithful mapping can be made, evolution of the laboratory system will mimic exactly that of the system of interest.  Moreover, very fine control of system parameters may be available in the laboratory system, enabling exploration of parts of phase space which are completely unavailable in the original system.   Such emulation of an inaccessible quantum system by a more controllable laboratory system, quantum simulation, may be one of the earliest attainable applications of quantum information processing (QIP), as a collection of tens of interacting qubits may already provide a significant speedup when compared to a classical computer.

Trapped ions are one of the most advanced schemes for QIP, with clear demonstrations of high-fidelity state preparation, one-qubit rotations, two-qubit interactions, and state determination having been achieved in the last several years~\cite{wineland98a,schmidtkaler03a,leibfried03a,haljan05a,acton06a}.  Along with plausible suggestions for scaling the system to many qubits~\cite{kielpinski02a}, this makes trapped ions an attractive system for the implementation of large-scale quantum simulation.  Proposals for quantum simulations of interacting spin systems~\cite{porras04a}, analogs to the Unruh effect and particle production in an expanding universe~\cite{alsing05a}, and effects analogous to the behavior of Dirac particles~\cite{lamata07a} demonstrate the wide range of quantum systems which may be emulated using the rich dynamics available in collections of trapped ions addressed with electromagnetic radiation.

A possible drawback is that due to the use of laser radiation to bring about interactions between ion qubits, many-qubit simulations may be hindered by the eventual spontaneous scattering events which decohere evolving many-body states.  The high rate of this scattering is due to the short natural lifetimes of intermediate atomic levels.  This is compounded by the lack of a straightforward decoherence-free subspace method, or quantum error-correction scheme in a fault-tolerant implementation, for quantum simulations with hyperfine (or optical) qubits in atomic systems~\cite{brown06b,brown07a}.  With scattering rates of one scattering event per hundred gatelike operations (consistent with present experimental constraints), and the assumption that simulation durations will be on the order of several gate operations, errorless simulations will be limited to a couple of tens of ion qubits.  Performing stimulated-Raman excitation at larger detunings from fast-decaying levels than currently used~\cite{wineland03a} can reduce the scattering rate by a few orders of magnitude~\cite{ozeri05a,ozeri07a}, though at the cost of greatly increased optical power requirements.  Nonetheless, even with gate errors near the so-called fault-tolerant threshold for quantum computing (an error probability of 10$^{-6}$ -- 10$^{-2}$ per gate operation is the range of recent estimates~\cite{aharonov97a,preskill97a,steane03a,knill05a,svore05a}; we will take 10$^{-4}$ as a reasonable value for calculation), quantum simulations would be limited to a couple of thousand qubits.  While this is certainly sufficiently large when compared to the capabilities of classical computers, it would be more reassuring to at least be able to glimpse a plausible route to very large (i.e., approaching mesoscopic scale) systems.

Here we describe a method for drastically reducing the error from spontaneous scattering and spontaneous emission during trapped-ion quantum simulation by using microwave (MW) and radio-frequency (RF) fields created by subwavelength structures to bring about system evolution.  In addition, the architecture we describe may be amenable to one-way cluster-state quantum computing~\cite{raussendorf01a} with the same reduction in error probability due to spontaneous scattering during cluster-state formation.  We note that the basic method of bringing about spin-spin interactions could also be used to perform two-qubit phase gates with an error rate below the fault-tolerant threshold, as the primary limiting fundamental error in the highest-fidelity demonstrations to date has been due to spontaneous scattering~\cite{leibfried03a}. The layout of the paper is as follows.  Section~\ref{sec_array} describes the basic experimental implementation of the array and ion states to be used, and in Secs. \ref{sec_interactions} and \ref{sec_scaling} we explore Hamiltonian interactions that can be brought about with this method, their attainable strength with reasonable experimental assumptions, and their scaling behavior as the trapping structure size is varied.  We describe some practical considerations for implementing simulations in one-dimensional arrays and ladders in Sec.~\ref{sec_applications}. In Sec.~\ref{sec_extensions} a plan for the use of this architecture for one-way quantum computing with trapped ions is described, and in Sec.~\ref{sec_discussion} we discuss the implications and limitations of this architecture, as well as giving a comparison with other methods.

\section{Microtrap and gradient coil array\label{sec_array}}

Traditional macroscopic RF Paul ion traps~\cite{paul90a} do not easily lend themselves to the formation of large-scale ion arrays.  Linear traps allow for containment of one-dimensional strings of ions, but for more than three ions in a single trap, the inter-ion spacing is a function of ion position in the trap~\cite{james98a}.  This makes a generalized ion-ion interaction tedious (see, e.g., \cite{leibfried05a}), if not impossible, to arrange.  In addition, the vibrational mode structure of many ions in a single trap is prohibitively complicated for sub-Doppler cooling, a necessity for processor initialization in almost all practical schemes.

The recent development of microfabricated ion trap structures can provide a solution to these problems through the separate trapping of individual ions in a collection of individual trapping zones that can be arrayed to suit a particular application, as suggested for large-scale QIP~\cite{wineland98a,kielpinski02a}.  In particular, surface-electrode traps~\cite{chiaverini05a,seidelin06a,pearson06a,labaziewicz07a} have the potential to create large one- and two-dimensional arrays of ions above the patterned surface of a trap chip, similarly to the trapping of neutral atoms  above atom chips~\cite{folman00a}.

\begin{figure}
\includegraphics[width=0.99 \columnwidth]{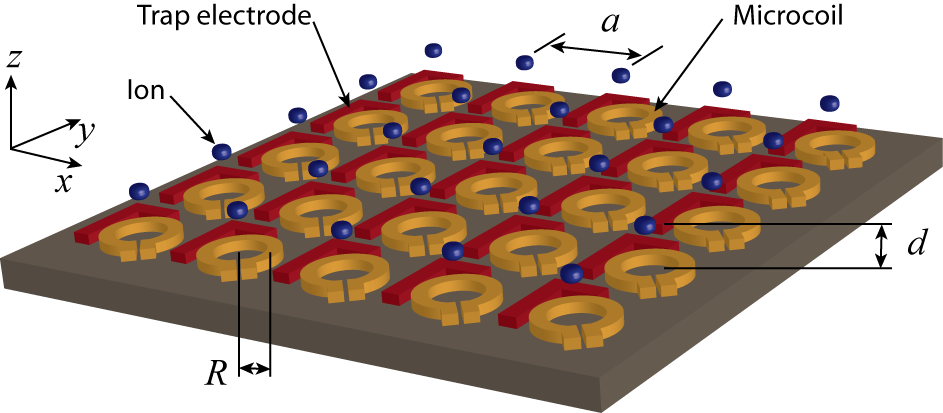}
\caption{(Color online) Rendering of section of surface trap array and microcoils.  Leads are not shown, and the electrodes shown are only a representative subset of those required.  Ions may be trapped directly above loops, or trap electrodes and coils may be offset so ions can be trapped above any point in a microcoil unit cell.  For field calculations, coils are treated as complete circular current loops.\label{fig_trap_array}}
\end{figure}

Although an arbitrary lattice layout, including ladders, frustration-prone geometries, etc., may be constructed (see Sec.~\ref{sec_applications}), we imagine here a one- or two-dimensional (1D or 2D) lattice of sites.  To be general, we will first discuss a 2D square array (see Fig.~\ref{fig_trap_array}). An ion is trapped at each site in its own electromagnetic microtrap potential formed by local surface-trap electrodes.  At each site, among the trapping electrodes, a metal coil is patterned on the chip surface; the coil may or may not be directly under the locally trapped ion.  The coil will provide excitation of the ion(s) above it through application of a MW or RF field through on-chip (or through-chip) transmission lines to the coil.  This structure will be of a size (radius $R$) comparable to the chip-ion distance $d$, and somewhat smaller than the resting inter-ion distance $a$.  This will bring about effectively near-field excitation of the hyperfine states of the ion as $R$ and $d$ will be on the order of 5~$\mu$m, whereas the wavelength of the relevant radiation (resonant with megahertz trap frequencies and gigahertz qubit splitting) is greater than $\sim\nobreak10$~mm.  The magnetic field induced by the displacement current above the coils is negligible; the ions are in the near zone where the fields, though oscillatory, are static in character, and this induced field is comparable to the directly created field only in the radiation zone (i.e., for points much further from the source than the wavelength)~\cite{jackson99a}.  This architecture has the advantages that (i) there will be a large gradient of the electromagnetic field of the coil at the trapping site, and (ii) the field amplitude can be made to drop off quickly with the distance from the site toward neighboring sites if desired.

Condition (i) is necessary for coupling to the motion of trapped ions.  An optical dipole force, typically generated by a standing (or more typically ``walking'') wave optical field, can exert a force on a strongly bound ion, because its wavelength is comparable to the extent of the ion's trap-state wave function (usually tens of nanometers).  This is not the case for MW or RF radiation, as there is no appreciable gradient in the field strength across the ion's wave function.  The gradient will in this case be brought about by the small extent of the wire structure used to create the electromagnetic fields.

Condition (ii) allows addressing of individual sites.  While this should not be necessary for many types of quantum simulation, as most of the different interactions can be employed uniformly, this is a desirable quality for other applications.  As an example, such site-specific addressing would allow for one-way quantum computing in such an architecture.  After cluster-state initialization, one-way quantum computing requires individual-site rotations and measurements which occur in a specific time ordering (this is described in more detail in Sec.~\ref{sec_extensions} below).

There have been suggestions for coupling internal states to ion motion using global magnetic field gradients~\cite{mintert01a}, but these proposals required many ions in one trap with a linear gradient over a large array, and no prescription for how to create the gradient was suggested.  In addition, these schemes would suffer from a susceptibility to variation in space in the gradient, due to the relatively large scale over which it must be maintained, and no description of multi-ion interactions was suggested.  Our current proposal links the trap structure to the gradient source conductors, registering the ion directly to the field of the chip; hence gradient spatial variation is much less likely.  In this way it is similar to a recent proposal to perform quantum operations using a static, spatially varying magnetic field produced by permanent magnets on a surface, through which ions are transported~\cite{leibfried07a}.  In our case ion motion, and the requisite velocity control, is not required, and the interactions can be controlled locally throughout an array on a potentially much faster (electronic) time scale.  Below we describe the use of the switchable microcoil-produced gradients to bring about multiqubit interactions suitable for quantum simulation or quantum computing.

\section{Engineering qubit interactions\label{sec_interactions}}

For simplicity, we consider the array described above in one of its two directions of smallest lattice constant $a$.  Along this direction, the electromagnetic field due to the array of microcoils will be periodic with spatial frequency 1/$a$.  However, the field magnitude can vary on a length scale in this direction smaller than the lattice constant, due to the smaller structure of the coils.  Hence large field gradients are possible; this is the basis of the state-dependent forces required to bring about ion-ion internal-state interactions.

For internal states whose energy varies with the magnetic field, a field gradient will lead to a force.  If this gradient is state dependent, a state-dependent force is created.  This leads to excitation of the ions' shared vibrational
modes in a manner dependent upon the internal states of the ions.  This can be made equivalent to an ion-ion interaction or, in quantum-information language, a multiqubit quantum gate.

\begin{figure}

\mbox{a) } 
\includegraphics[width=0.92 \columnwidth]{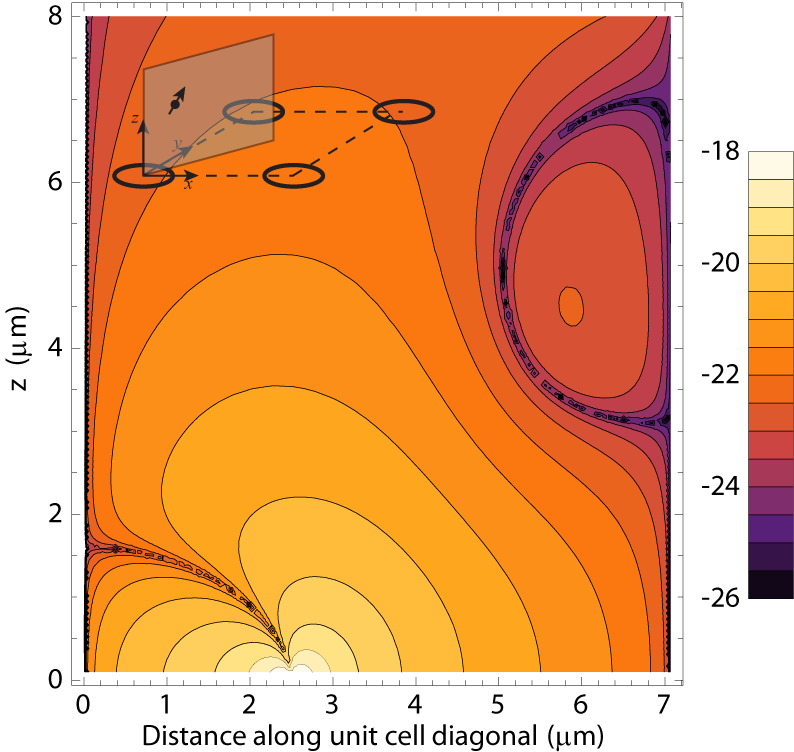}\\
\mbox{b) }
\includegraphics[width=0.85 \columnwidth]{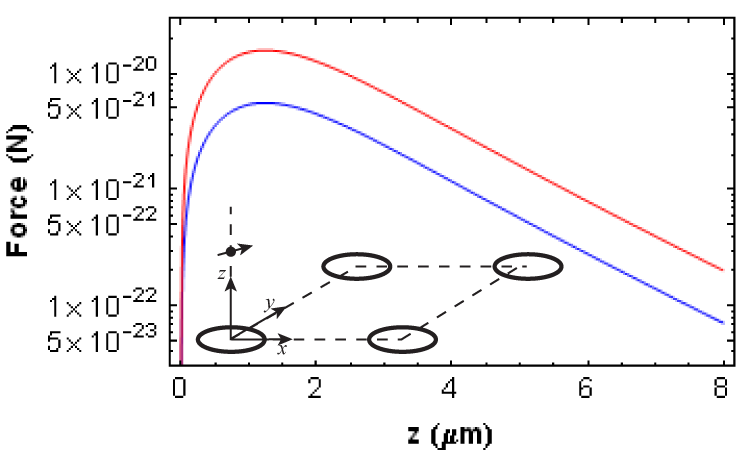}

\caption{(Color online) Force on an ion above an array microtrap.  These data are for positions near the center of a $20\times20$ array.  (a) Logarithm of the absolute value of the $x$ (or $y$) component of the force in newtons on an ion produced by an array of microcoils.  The lattice spacing is 10~$\mu$m, with a current amplitude of 10~mA in each loop of radius 2.5~$\mu$m.  The horizontal axis is distance along the unit cell diagonal, and the vertical axis is the ion's height above the trap-microcoil surface; the relative geometry of a moment in a general location in this plane is shown in the inset diagram.  The center of the nearest microcoil is located at 0 on this axis.  See Sec.~\ref{sec_scaling} for spin-spin interaction strength attainable with these forces. (b) Force on a moment when its orientation is pinned by an external field.  The lower (upper) curve is the logarithm of the $x$~component ($z$ component) of the force on an ion as a function of ion height above the surface of an array with ions centered on loops in the $x$ and $y$ directions and the ion moment pinned in the horizontal $x=y$  (vertical $z$) direction by an external dc magnetic field; the inset diagram shows the relative geometry of a moment in a general position along the axis. \label{fig_force}}
\end{figure}

The force on a dipole {\bf m} (the ion in various of its Zeeman states) due to a magnetic field {\bf B} (the field from the microcoil) is ${\bf F}={\bf \nabla}({\bf m}\cdot{\bf B}$).  We will assume that the direction of each ion's dipole adiabatically follows the local field at all times; this will be true as long as the RF frequency is significantly less than the Larmor precession frequency of the qubit states in the field, a condition that will be met for attainable magnetic field amplitudes (in the gauss range).  We first consider a force in the $x$-$y$ plane ($z$ is normal to the trap surface).  As the component of the moment along the field remains constant, the magnitude of the force in the $x$ direction reduces to $F_{x}=m_{x}\frac{\partial B_{x}}{\partial x}+m_{y}\frac{\partial B_{y}}{\partial x}+m_{z}\frac{\partial B_{z}}{\partial x}$.  Above the center of one of the microcoils, the dipole will be aligned with $z$, and there will be no force in the $x$ (or $y$) direction.  However, if the dipole is situated along the $x=y$ diagonal slightly off the symmetry axis of the microcoil, there will be a significant force component along $x$ and $y$, $F_{x}=F_{y}$, with a small $z$ component as well.  Figure~\ref{fig_force}(a) is a contour plot of the logarithm of the $x$ (or $y$) component of the amplitude of the force produced by a loop array with $R=2.5$~$\mu$m, $a=10$~$\mu$m, and a current of 10~mA as a function of  distance along the unit cell diagonal from the center of one loop and distance above the chip surface.  Continuous currents of this magnitude or greater are attainable in conductors of a few square micrometers cross section in vacuum at room temperature~\cite{groth04a}.  Points approximately above the coil perimeter are best for producing large forces.  If the ion dipole always points in the direction of the magnetic field, a RF current applied to the loops at frequency $f_{\rm RF}$ will produce a force at frequency 2$f_{\rm RF}$, since at both field extrema the force will be in the same direction, going to zero at the field zero-crossing points of a RF cycle.  That is, the direction of both the moment and the field gradient will change sign on going from the positive RF extremum to the negative, leading to a force in the same direction at the extrema.

If we assume that an overall constant magnetic field, somewhat larger than the amplitude of the oscillating field, is applied in addition to the field from the coils, we can approximately pin the ion magnetic moment in a particular direction, and the field gradient is unaffected.  This allows more freedom in positioning the ion relative to the microcoils for a lateral ($x$ or $y$) force.  It is also typically useful to work in a small constant field to split degenerate Zeeman sublevels.  If we align the moment with the $x=y$ direction, we can get a force for ions positioned directly above each microcoil.  In this case the force component in the $x$ (or equivalently $y$) direction is $F_{x}=\frac{1}{\sqrt{2}} m\left(\frac{\partial B_{x}}{\partial x}+\frac{\partial B_{y}}{\partial x}\right)$.  With the ions trapped directly above the coil centers, this geometry has the advantage of greater symmetry and hence may be easier to fabricate.  The force on an ion in the $x$ direction as a function of height above the coil for this configuration is shown in Fig.~\ref{fig_force}(b) (lower curve).  In this case the force on the ion will be at a frequency $f_{\rm RF}$ as the forces at the RF field extrema will be in opposite directions.

The above calculations are for force in the $x$ or $y$ (lateral) direction, but a force in the $z$ direction can excite the vertical modes and bring about a different interaction (see below).  For a moment aligned with the $z$ direction and positioned on the axis directly above a coil, there will be a significant force $F_{z}= m\frac{\partial B_{z}}{\partial z}$ in the $z$ direction as the gradient in this direction is sizable.  Figure~\ref{fig_force}(b) (upper curve) shows this force as a function of height above the loop.

For the trapped-ion moment, we consider each two-level quantum system (analogous to a spin 1/2 or a qubit) being composed of two sublevels of the ground-state hyperfine manifold.  The hyperfine splitting for ions of interest is typically a few to several gigahertz, and in a small overall quantizing magnetic field, each of the hyperfine levels obtains a Zeeman substructure on the megahertz scale.  It is possible to use two of these sublevels, (i) one in each hyperfine level, or (ii) two in the same hyperfine level.  In the former case, the so-called clock states ($|F, m_{F}=0\rangle$ and $|F', m_{F}'=0\rangle$) may not be used, as there is no differential force dependent on qubit state.  For many simulations, a single-qubit rotation (such as would be brought about by a global simulated field) is desired, simultaneous with the spin-spin interaction.  Thus it may be beneficial for the qubit levels to be defined as in case (i), with the additional stipulation $|m_{F}-m_{F}'|\leq1$ so that magnetic dipole transitions can enact the global interaction.  We will consider this case, with the two qubit levels, defined as $|\!\uparrow\rangle\equiv|F,m_{F}=1\rangle$ and $|\!\downarrow\rangle\equiv|F',m_{F}'=0\rangle$ split by $\omega_{0}$.  In ions with nuclear spins greater than 1/2, other states with larger $z$ spin projections will enable larger forces, but we consider the states defined above as a general reference point.  Similarly, larger forces may be obtained using states as in case (ii) due to the larger gyromagnetic ratios available in the ``stretched'' states (states with the largest value of $|m_{F}|$ in a particular hyperfine level).  It should also be pointed out that first-order magnetic-field-independent states~\cite{langer05a} cannot be used for the qubit levels, though in an application where the interaction is not continuous for the duration of the protocol,  e.g., cluster-state computation, the high-fidelity single-bit rotations provided by MW excitation would allow state transfer into and out of the protected manifold between operations (cf. Sec.\ref{sec_extensions}). 

Each ion in the array will be situated above (though not necessarily above the center of) a microcoil in its own microtrap.  Through application of a RF signal near the trap frequency to the microcoils, a state-dependent force is created that excites collective motional states.  The motion leads to the acquisition of a geometric phase dependent on the internal ion states, and hence a spin-spin interaction.  The two qubit levels behave differently as a function of magnetic field.  To first order, the $|\!\downarrow\rangle$ state does not change as a function of the field, whereas the $|\!\uparrow\rangle$ state acquires a linear Zeeman shift.  Hence ions in the $|\!\downarrow\rangle$ state will not feel a force due to the field gradient, but ions in the $|\!\uparrow\rangle$ state will.  The internal state of a pair of ions in a general superposition will in general be entangled with the ions' shared motional state due to such an interaction:  different parts of the superposition will feel different forces and thus evolve to different motional states.

If the state-dependent force is applied slightly off resonance with one of the ions' motional modes, certain parts of the superposition can pick up geometric phases relative to the others as a function of time; this is the mechanism behind the two-ion-qubit phase gates used for circuit quantum computation~\cite{leibfried03a,haljan05a}.  This state-dependent phase acts exactly like a spin-spin interaction.  For instance if the two-ion states $|\!\uparrow\downarrow\rangle$ and $|\!\downarrow\uparrow\rangle$ acquire a phase $\Phi_{\rm ZZ}(t)$ relative to the states $|\!\downarrow\downarrow\rangle$ or $|\!\uparrow\uparrow\rangle$, we have a situation analogous to the evolution of two-spin system under a Hamiltonian such as

\begin{equation}
H_{I}= J \sigma^{z}_{1} \sigma^{z}_{2}.
\end{equation} 

\noindent  For this interaction, the asymmetric spin states of a superposition will acquire a phase $\Phi_{I}(t)=e^{-2iJt}$ relative to the symmetric spin states.  It should be noted that for typical implementation of the phase gate as described above, $\Phi_{\rm ZZ}$ is not generally an imaginary exponential linear in $t$, but linearity is a good approximation for sufficiently large detunings from the motional-mode resonance compared to the inverse total evolution or interaction time~\cite{leibfried03a}.

For the array structure defined above, this type of interaction may be shared by all the ions in the array, but the interaction can be made local.  In the case of microtraps, with the Coulomb energy between neighboring ions much weaker than the trapping potential of each trap, the ions' internal spin states will have a dipolarlike interaction~\cite{porras04a,deng05a} with a strength proportional to $1/b^3$ ($b$ is the distance between two arbitrary ions).  Hence nearest-neighbor interactions may be approximated.  This situation can be exploited to implement large-scale simulations of interacting-spin Hamiltonians.  A ferromagnetic (negative $J$) spin-spin interaction may be brought about using the lateral modes, while an antiferromagnetic (positive $J$) interaction may be produced using the vertical modes~\cite{porras04a}.

Global field terms may be implemented by sending a MW signal resonant with the qubit frequency to all the coils.  This can bring about single-spin evolution like that due to a Hamiltonian such as

\begin{equation}
H_{G}= B_{f} \sigma^{x}.\label{eq_bfield}
\end{equation}

\noindent Thus Rabi oscillations of the ions' internal states mimic exactly the precession of spins in a fictitious transverse magnetic field $B_{f}$ (here in energy units).  As the MW is produced by all coils simultaneously, a field of uniform strength will be present at all ion positions in the bulk of the array, with a small variation for ions at the array edge (periodic boundary conditions may be arranged, however, to remove these edge effects in some cases---see Sec.~\ref{sec_applications}).  This type of interaction may also be brought about by a separate large MW coil outside the array extents as only a uniform MW field is required.  This latter method may prove simpler than mixing signals to the microcoils, and it has the important advantage of allowing the evolution to be totally independent of the ions' motional state, for the same reason that fields with large gradients are required to excite the motional states.  Note that MW excitation is not subject to the Larmor precession frequency criterion mentioned above for RF excitation, because we desire that the MW induce spin flips and must be at the Larmor frequency corresponding to $\omega_0$.

Combining the interaction $H_{I}$ and global-rotation $H_{G}$ Hamiltonians above, and generalizing to an array of spins, we get the Ising model in a transverse magnetic field:

\begin{equation}
H_{\rm Ising}= \sum_{\langle i,j \rangle} J_{ij} \sigma^{z}_i \sigma^{z}_j + B_{f} \sum_{j} \sigma^{x}_{j}.
\label{eq_ising}
\end{equation}

\noindent Here $J_{ij}$ is the Ising interaction strength between ions $i$ and $j$ and $B_{f}$ is the global transverse magnetic field interaction strength.  This model is exactly solvable in 1D, though not generally for dimension 2 or higher.  By combining the two experimental procedures described above for ions in a microtrap array, one can bring about evolution that emulates evolution under Eq.~(\ref{eq_ising}).

This evolution will make a good test case to demonstrate that the basic interactions can be emulated efficiently and accurately. It will be useful to verify the behavior as a function of the interaction strength ratio $\gamma=J/B_{f}$, especially in the limits $|\gamma|\ll 1$, $|\gamma|\gg 1$, and the interesting region $|\gamma|\approx 1$.  Here the competition between local interaction and a global field is the most pronounced, and this is the regime where a quantum phase transition would occur for a system of many spins.  With success in calculable cases (for small numbers of spins, many spins in 1D, etc.), more complicated interacting many-spin Hamiltonians may be explored in 1D and 2D.  

Many extensions of the above model may be explored.  Heisenberg-like models will be possible with the introduction of another similar interaction, in combination with the Z-Z interaction, but along another spin axis.  Almost arbitrary array layout is possible with microfabrication of arrays as described here, and individual-site addressability will allow for defect placement, pinning, or site-specific tailored spin behavior.  For instance, the microtrap array can allow for the inclusion of defects in a simulation, since ions are trapped individually.  Where a particular ion is absent, or where a specific microtrap's frequency is altered, local interactions will be affected.  The possibility of site-specific addressing through individual coils can provide the opportunity for pinning, as well.  An ion in a particular site can be strongly controlled from the local coil via a MW signal that acts like a magnetic field in a particular direction [as in Eq.~(\ref{eq_bfield})]; these pinned sites can be time dependent as well, since the MW switching can be controlled on a fast time scale.  The phase or the amplitude of the MW delivered to a particular site may be altered to change the direction or magnitude of the field locally; in effect, the $B_{f}$ in Eq.~(\ref{eq_ising}) can be made to be a function of the index $j$.  Defects or pinned sites can be placed regularly or at random throughout the array.  The addressability of the array described above is not perfect, however, and all local modifications will have an inverse cubic (or stronger) scaling with distance.  Hence there will be nonzero evolution at neighboring sites.

\section{Strength and scaling properties of the interactions\label{sec_scaling}}

\begin{figure}

\mbox{a) } 
\includegraphics[width=0.92 \columnwidth]{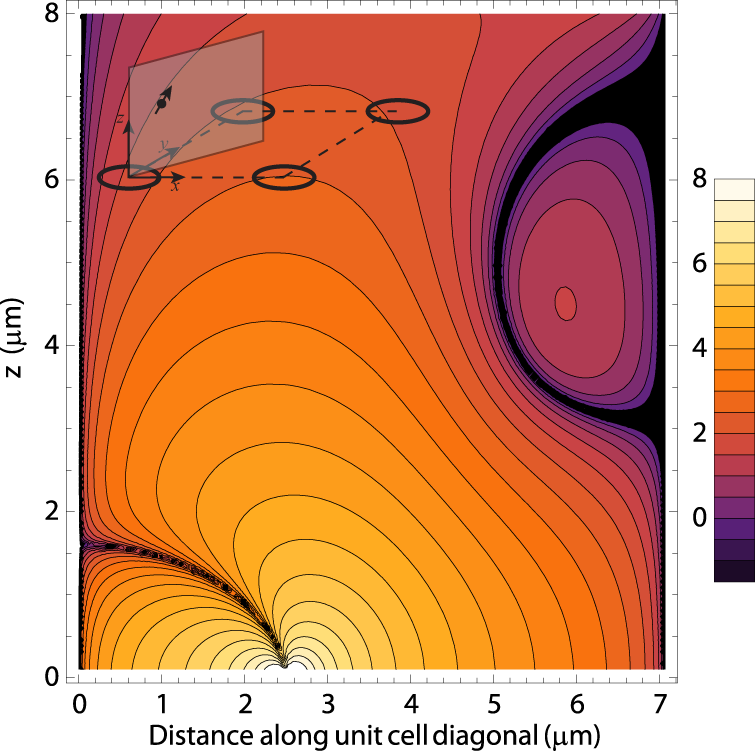}\\
\mbox{b) } 
\includegraphics[width=0.85 \columnwidth]{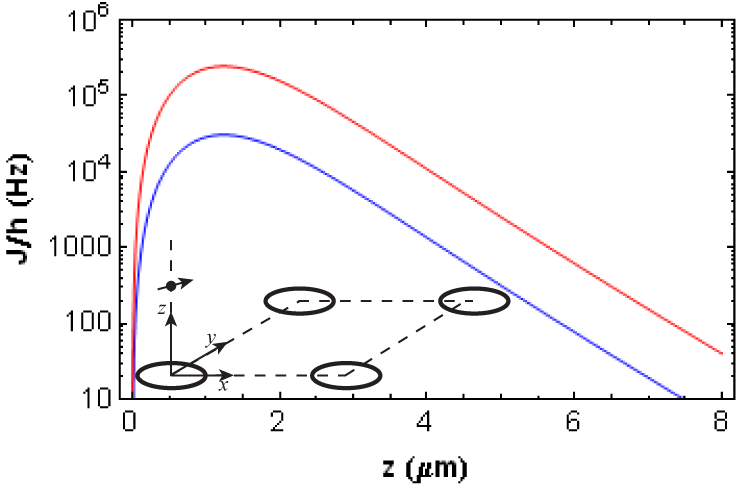}
\caption{(Color online) Interaction strength $|J|$ for the same configuration as in Fig.~\ref{fig_force} above. (a) Logarithm of the nearest-neighbor interaction strength (in hertz) produced by a $20\times20$ array of microcoils.  The lattice spacing is 10~$\mu$m, with a current amplitude of 10~mA in each loop of radius 2.5~$\mu$m.  The horizontal axis is distance along the unit cell diagonal, and the vertical axis is the ion's height above the trap-microcoil surface (see inset diagram).  The center of the nearest microcoil is located at 0 on this axis, and an ion of mass of 9~amu ($^9$Be$^+$) is assumed.  Interactions in the 1--100~kHz range are attainable at locations 1--5~$\mu$m above the coil edge (vertically above 2.5~$\mu$m on the horizontal axis). (b) The lower (upper) curve is the logarithm of nearest-neighbor interaction strength as a function of ion height above the surface of an array with ions centered on loops in the $x$ and $y$ directions and the ion moment pinned in the $x=y$ direction ($z$ direction) by an external dc magnetic field (see inset diagram).\label{fig_inter}}
\end{figure}

The interaction strength for neighboring sites, in the case where the microtraps are stronger than the inter-ion Coulomb interaction (strong binding), is given by~\cite{porras04a,deng05a}

\begin{equation}
J=\alpha \frac{1}{4\pi\epsilon_0} \frac{F^2 e^2}{m^2 \omega_{\rm T}^4 a^3},\label{eq_J}
\end{equation}

\noindent the dipolar interaction mentioned above.  Here $F$ is the force on $|\!\uparrow\rangle$ due to the magnetic field gradient, $e$ is the ion charge, $m$ its mass, $\omega_{\rm T}$ is the trap frequency in radians per second, and $\alpha$ is a signed constant of order 2.  The interaction strength is plotted in Fig.~\ref{fig_inter} for the same cases as in Fig.~\ref{fig_force}, and the interaction strength for a few different species of ion is listed in Table~\ref{tab_calcs}.

Equation~(\ref{eq_J}) is valid for strong binding, i.e., where the quantity $\beta$ satisfies

\begin{equation}
\beta \equiv \frac{1}{4\pi\epsilon_0} \frac{e^2}{m \omega_{\rm T}^2 a^3}\ll 1.\label{eq_binding}
\end{equation} 

\noindent This criterion is difficult to satisfy while symmetrically shrinking the trap and coil geometry.  The trap frequency is determined by the trap size $\rho$ (the distance from the ion to the nearest trap electrode) and voltage $V$ as $\sqrt{V}/\rho$ for the lateral modes, and with the consideration of limiting the local field below the breakdown value, $V\propto \rho$ and $\omega_{\rm T}\propto \rho^{-1/2}$.  The left-hand side of the inequality in Eq.~(\ref{eq_binding}) will vary as $\rho/a^{3}$.  If we restrict the geometry such that this criterion is met as we scale down the trap, $a$ must scale as $\rho^{1/3}$. 

\begin{table*}
\caption{Spin-spin interaction strength for ions above a microcoil array for the case where the moment follows the magnetic field, here tabulated for a few light ion species at chosen trap frequencies.  Each ion is held above the edge of a microcoil along the $x=y$ direction of the array at a height given by $d$.  The array spacing $a$ is fixed at 10~$\mu$m.  The nuclear spin $I_{N}$ is listed, as there is the possibility to use states with higher values of $|m_F|$ to obtain larger forces ($|J|$ values listed here are for $|m_F|=1$ as described in the text).  In this case, the interaction strengths listed should be multiplied by $|m_F|^2$ for the particular level used.\label{tab_calcs}}
\begin{ruledtabular}

\begin{tabular}{l c r r r r r r r}
Ion  &  $I_{N}$ & $\omega_0/2\pi$  &  $\omega_{\rm T}/2\pi$ & $\beta$ &  \multicolumn{3}{c}{$|J|$ (kHz)}\\
     &      &           (GHz)  &                 (MHz)  &         &  $d$=1~$\mu$m & $d$=2~$\mu$m & $d$=5~$\mu$m\\
\hline
$^{9}$Be$^+$  & 3/2 & 1.25 &  1.00 & 0.38 &  160 & 15  & 0.12\\
$^{25}$Mg$^+$ & 5/2 & 1.79 &  0.75 & 0.25 &  84 & 7.9  & 0.063\\
$^{43}$Ca$^+$ & 7/2 & 3.23 &  0.60 & 0.22 &  53 & 4.9  & 0.039\\

\end{tabular}
\end{ruledtabular}
\end{table*}

The state-dependent force is determined by the microcoil size $R$ and chip-ion distance $d$.  As noted above, the trap frequency will go as $\rho^{-1/2}$, and in the near field the force (through the magnetic field gradient) will vary as $1/r^2$ for $R,d \propto r$ and constant microcoil current.  Hence the interaction will scale as

\begin{equation}
|J| \propto \frac{F^2}{\omega_{\rm T}^{4} a^{3}} \propto \frac{\left(r^{-2}\right)^{2}}{\rho^{-2} \rho} \propto r^{-4} \rho.
\end{equation}

\noindent If we make all these lengths scale together ($R,d,\rho \propto r$), the interaction will scale as $r^{-3}$ as the inter-ion distance is reduced.  We will need to scale the current in the microcoil down as well, however, to avoid exceeding the critical current density as the coil shrinks.  If we scale the current as $I\propto r^2$ to maintain a constant current density in the coil, the force will scale as a constant, so that the strength of the interaction will vary as

\begin{equation}
|J| \propto \frac{1}{\omega_{\rm T}^{4} a^{3}} \propto \frac{1}{\rho^{-2} \rho} \propto r,
\end{equation}

\noindent enabling gains in the interaction when the entire architecture is expanded.  However, upward scaling is not tenable as the inter-ion distance must grow more slowly than the system size, $a\propto r^{1/3}$, as determined above.  The microcoils will eventually overlap and the scaling will break down.  Using vertical ion vibrational modes, the frequencies also will scale as $\rho^{-1/2}$, though by slightly different arguments relating to RF Paul trap stability~\cite{chiaverini05a}.  Thus $a$ would have to scale as $r^{1/3}$ in this case as well, leading to the same conclusions for scaling.

Global-field-type interactions like Eq.~(\ref{eq_bfield}) will obviously not scale with the architecture for global-field interactions implemented with a separate larger MW coil outside the array.  However, if implemented with the on-chip coils, the Rabi frequency of rotations due to such an interaction will be proportional to the microcoil magnetic field strength at the ion position, which will vary as $I/r$.  If we scale the current with $r^2$ as described above, the Rabi frequency will scale as $r$.

Though the scaling of these interactions does not appear to benefit from reduced trap size, 1--10~$\mu$m scale structures can produce spin-spin interactions in the 1--100~kHz range (global interactions will be up to the megahertz range) for reasonable coil current magnitudes.  With the reduction in spontaneous scattering probability inherent in this scheme, the speed of these interactions will be sufficient for significant system evolution in an experimentally feasible time period, even for geometries near the larger end of this scale with slower interactions.  Larger systems may benefit from on-chip delivery of long-wavelength field excitation of global (one-qubit) rotations, though the trap frequencies decrease as system size increases, making initial cooling of the ions' vibrational states increasingly difficult.

\section{Selected simulation applications in one dimension \label{sec_applications}}

As alluded to in Sec.~\ref{sec_interactions}, the proposed architecture allows for an almost arbitrary layout of an ion lattice, making technologically viable several 1D problems that have recently been of great interest.  The specific geometries considered in this section are single rings (a 1D array wrapped around on itself) and two concentric rings (spin ladders such as Heisenberg ladders).

\subsection{Ring geometry}

After successful demonstration of a quantum simulation of the elementary interaction described above ($H_{\rm Ising}$) in a small collection of spins, the next step will be to extend this technique to much larger systems of ions to directly attack problems intractable by other means, such as long chains of interacting spins.  With the proposed architecture one could easily make a long straight-line chain of spins, but a ring layout solves several problems.  One advantage of the ring geometry is that for a given overall ``trap-chip'' size, one can fit significantly more ions into a ring than in other symmetric configurations.  For example, with the lattice spacing $a$ set to 10~$\mu$m, a straight configuration would be limited to 300 ions in a 3-mm-long area, as opposed to 2000 ions in a 3-mm-diameter ring.  A major advantage is the elimination of undesired boundary conditions.  Edge effects at the ends of a linear chain could have a large confounding effect on simulation outcome, such that many ions at each end would have to be discarded in experimental analysis.  With a ring configuration one obtains the ability to simulate 1D spin systems with periodic boundary conditions, and the edge effects can be eliminated.  Thus useful data may be obtained from the entire array while allowing for investigations of phenomena such as spin waves~\cite{deng05a} without complications like reflections or dispersion near the array end points.

\begin{figure}
\includegraphics[width=0.80 \columnwidth]{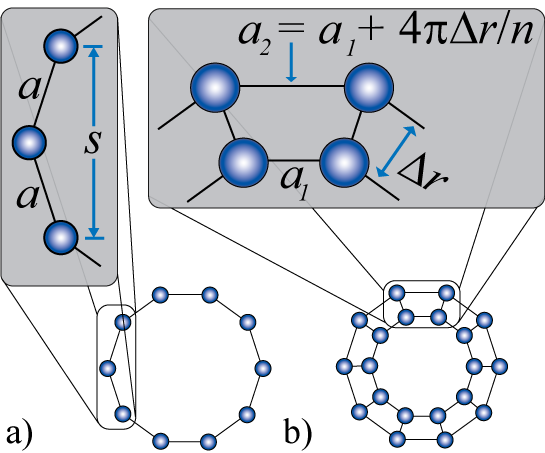}
\caption{(Color online) Ring and ladder geometries with periodic boundary conditions. (a)  1D ring:  the nearest-neighbor distance is $a$ and the next-nearest neighbors are separated by a distance $s$ dependent on the number of ions in the ring.  The NNN interaction strength can be made to closely approach that in a straight-line array for tens of ions. (b) Ladder with periodic boundary conditions:  the interaction strength along the exterior leg (length $a_2$) approaches the strength along the inner leg (length $a_1$) within a few percent for a few hundred pairs of ions.  The rung length $\Delta r$ is the difference in the radii of the inner and outer rings, and $n$ is the number of rungs. \label{fig_ringgeom}}
	
\end{figure}

For the purposes of 1D Ising-like simulations, it is desirable to minimize the error introduced in next-nearest-neighbor interactions due to the ring geometry.  Referring to Fig.~\ref{fig_ringgeom}(a) and Eq.~(\ref{eq_J}), we see that the ratio of next-nearest-neighbor (NNN) interaction strength to nearest-neighbor (NN) interaction strength [$J(s)/J(a)$] is slightly larger than it would be for a linear chain.  Simple geometric considerations yield

\begin{equation}
s=a\left[2+2\cos\left(\frac{2\pi}{n}\right)\right]^{1/2}.
\end{equation}

\noindent Here $n$ is the number of ions in the ring.  It follows that, for more than 12 ions in a ring, the NNN to NN interaction ratio is less than 10\% bigger than for a straight-line array, and for more than 38 ions, the ratio is less than 1\% bigger.  At this level, we could say the ions don't ``know'' that they are not in a straight-line array, when considering experimental imperfections.  The ratio $J(s)/J(a)$ rapidly and asymptotically approaches $1/8$, and the slight increase in $J(s)/J(a)$ relative to the linear geometry can be made negligible for tens of ions.

\subsection{Ladder geometry}

There are various applications of spin ladders to low-dimensional problems in antiferromagnetism~\cite{barnes93a}, superconductivity, and complex entanglement~\cite{xin05a}.  For instance, hopping-boson and -fermion models can be mapped to interacting spin-ladder Hamiltonians via Jordan-Wigner-type transformations~\cite{batista01a}. Thus Hubbard-like models (with possible relevance to high-temperature  superconductivity) may be addressed by direct quantum simulation.

For some of these mappings, linear lattices with hopping particles map to interacting spins on ladders with different interaction strengths on the rungs and legs of the ladder~\cite{batista04a} including configurations with asymmetric ladders~\cite{chen03a}. This can be implemented in this architecture in one of three ways:  (i) by fixing the direction of the quantizing magnetic field that pins the ion moments such that it points not directly along the square array diagonal but along a direction whose $x$ and $y$ components are in proportion to the desired ladder-rung strength ratio (in cases with uniform lattice vectors throughout the array); (ii) by tailoring the geometry of the trap during fabrication such that the microtrap spacings in the ladder and rung direction are different, leading to different interaction strengths in those directions; or (iii) by modifying individual ion trap frequencies, possibly in combination with methods (i) or (ii), to locally modify the interaction [see Eq.~(\ref{eq_J})].

Ladder geometries will also benefit from the use of ring arrays with periodic boundary conditions.  While this architecture may incorporate ladders with more than two legs to explore complex materials systems~\cite{azzouz05a}, we focus here on concentric rings forming a two-leg ladder.  Referring to Fig.~\ref{fig_ringgeom}(b), it can be seen that a rectangular lattice cell of a straight-line ladder is transformed into a trapezoid.  The interaction strengths of ion pairs separated by inner and outer leg distances $a_{1}$ and $a_{2}$ are slightly different.  The length of the ladder rungs is $\Delta r=r_{2}-r_{1}$ where the ions are located at the vertices of regular polygons inscribed on concentric circles of radii $r_{1}$ and $r_{2}$.

So moving from the linear architecture to the ring architecture leaves the rung interaction strengths unaltered for a given leg separation $\Delta r$.  However, it introduces some asymmetry to the longitudinal ladder-leg interactions.  This may be desired for some applications as mentioned above.  It is, however, also necessary to have symmetric ladders in many cases.  The ladder legs differ in length such that

\begin{equation}
a_{2}= a_{1}+\frac{4\pi\Delta r}{n},
\end{equation}

\noindent where $n$ is now the number of rungs (half the number of sites).  Comparing the relative interaction strengths on the two legs, $J(a_{2})$ and $J(a_{1})$, we find that for $\sim\nobreak180$ or more rungs the interaction strengths of two legs differ by less than 10\% for $a_{1}= 2 \Delta r$, and for $\sim\nobreak1900$ or more they differ by less than 1\%.  For thousands of rungs the ladder ring looks essentially like a straight-line ladder, but even for just a few hundred rungs it should be possible to compensate for this asymmetry using the methods described above.  For a few to a few hundred rungs, asymmetric spin ladders may be implemented without much compensation; for instance, $J(a_{2})/J(a_{1})=0.5$ for $\sim\nobreak25$ rungs.

As it may be difficult to measure the internal state populations of many ions at once, the ion ring could be rotated by translating ions around the ring in unison and unidirectionally, much like a revolver, through variation of the potentials on the trapping electrodes as part of a measurement procedure.  One (or several) of the angular positions on the ring could be an imaging location, and an ion's state could be measured before revolving the next ion into place, a method for time multiplexing the measurements.  Another  related possibility for readout would allow spatially efficient simultaneous measurement:  the ring array would be transformed into a densely packed (as dense as allowed by the imaging optics resolution) 2D array by moving the ions via potential variation as above.  The array could be imaged onto an extended sensor [e.g., a charge-coupled device (CCD)], accomplishing simultaneous readout in minimal space~\cite{acton06a}.

\section{One-way cluster-state quantum computation\label{sec_extensions}}

Besides its application to quantum simulations, the architecture described here may form a natural layout for one-way quantum computing (OWQC) with cluster states of trapped ions.  Starting with a square array of ions, each near a microcoil as described above, each is prepared in an equal superposition state [e.g. $|+\rangle=\frac{1}{\sqrt{2}}(|\!\uparrow\rangle +|\!\downarrow\rangle)$\ ], which can be created from an optically pumped initial state using a single uniform MW pulse bringing about an $H_{G}$-type Hamiltonian for a fixed time.  A cluster state could then be created through application of the Z-Z interaction ($H_I$) for a small amount of time.  After this initialization, the computation consists only of classically controlled single-bit rotations (based on previous measurements) and subsequent measurements.  The rotations can be performed via MW fields at frequency $\omega_0$ created by the coil beneath each ion.  This built-in addressability makes laying out the quantum (one-way) circuit as easy as assigning each coil in a 2D array to a particular rotation.  The measurement will require a laser beam for resonance fluorescence, but if the ions are sufficiently far apart, the other qubits should not be disturbed.  For simplicity, the fluorescence could be read out through a nonpixelated detector if desired, as the time multiplexing inherent in the OWQC scheme avoids problems with simultaneous measurements.  A recent topological method for fault-tolerant OWQC in two dimensions~\cite{raussendorf07a} could be implemented in this type of system to provide a truly scalable path toward quantum computation with trapped ions.

\begin{figure}
\includegraphics[width=0.99 \columnwidth]{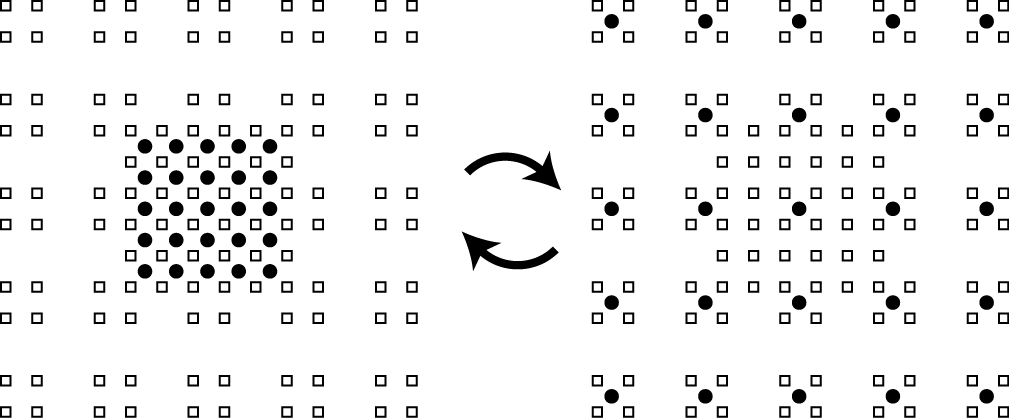}
\caption{Plan view depicting ion surface trap array with one possible electrode configuration and method for array expansion during one-way quantum computing protocol.  The ions ($\bullet$) are in a different plane from the electrode segments ($\Box$).  The left panel shows ions in a compressed array, each over a microcoil (not shown), in which configuration a cluster state could be formed as described in the text.  The ions would then be repositioned through concerted variation of the potentials on the electrodes to form an expanded array, as in the right panel, suitable for individual qubit rotations via laser or microcoil and readout via resonance fluorescence.  The ions would then be returned to the compressed array for the next experiment, or for the next stage in a topological error-correcting protocol~\cite{raussendorf07a}. \label{fig_owqc}}
\end{figure}

With the inclusion of the capability for controlled ion motion, the requirements for localized resonance-fluorescence beams can be relaxed somewhat.  Segmented trap electrodes have been used to shuttle ions between zones in rather complicated patterns~\cite{barrett04a,chiaverini04a,chiaverini05b} even to the point of reordering ions in a 1D configuration~\cite{hensinger06a}.  This technology can be naturally adapted for use in surface-electrode structures, as additional electrode segmentation requires only minor changes in the microfabrication process.  For a simplified-readout OWQC, the ions would start out in a 2D array with relatively small lattice constant such that the Z-Z interaction can be strong.  After initial cluster-state formation, the array can be expanded by a factor of 10 or so via ion motion caused by concerted variation of electrode segment potentials (see Fig.~\ref{fig_owqc}).  Then, all subsequent single-bit rotations and measurements may be carried out with large ion separations, such that the undesired overlap of resonance-fluorescence beam radiation will be negligible.  For instance, consider an initial $100\times100$ square array with 10~$\mu$m between ions in each direction.  After cluster-state initialization using RF excitation of all ions simultaneously, the ion configuration could be expanded to an array with 100~$\mu$m lattice constant in each direction.  Individual-qubit-rotation and resonance-fluorescence Gaussian laser beams with waists of 20~$\mu$m focused on a particular site would overlap nearest-neighbor ions with radiation intensity reduced by more than 20 orders of magnitude (scattered light from the chip would be the limiting factor in this case).  Individual rotations may also be done with MW from the microcoils to minimize spontaneous emission---the field amplitude at a neighboring ion will be less than 10$^{-3}$ times that at the target ion for $R=2.5$~$\mu$m.  An arrayed imaging system, such as a CCD as mentioned above, could be used to read out the ions' states, and the larger array size would allow for relatively simple imaging optics as high resolution or magnification would not be required (the whole array would be 1~cm on a side).  After readout, the ions could be shuttled back to the compressed array for the next calculation (starting with laser cooling and optical pumping) or next topological fault-tolerant error-correcting-code cycle.

Long OWQC computations may require many thousands of individual-bit rotations and measurements, so coherence of the ion-qubit ensemble is a clear consideration, and coherence times approaching a second or more may be required.  Recent progress in internal-ion-state coherence makes these times attainable, particularly with first-order magnetic-field-independent hyperfine qubits.  Coherence times of greater than 10~s have been demonstrated in these systems in a QIP setting~\cite{langer05a}, and coherence times almost two orders of magnitude longer can be obtained in some configurations~\cite{bollinger91a}.  As mentioned above, the ions cannot be in such protected manifolds during cluster-state formation using the method described here, but the qubits can be transferred from the $\{|\!\uparrow\rangle,|\!\downarrow\rangle\}$ space to the protected space after initialization and remain there for the bulk of the computation (individual rotations and measurements).  The cluster formation and state transfer procedures would require 10--100$~\mu$s and $\sim\nobreak$1~$\mu$s, respectively, and each can be performed on all ions in parallel.

The inverse cubic interaction strength produced with the Ising-like interaction in the described architecture will lead to imperfect cluster-state formation due to qubit-qubit couplings beyond the nearest-neighbor level.  It is interesting to note, however, that even with such extraneous long-range correlations, fault-tolerant computation is still possible if the correlations in a $D$-dimensional system drop off faster than $1/b^{D}$ for qubits a distance $b$ apart~\cite{aharonov06a}.  The inverse cubic falloff described above satisfies this condition for one- and two-dimensional arrays.  Additionally, imperfect cluster states, e.g., those created with correlated ``noise'' as is present here, may be distilled to higher-fidelity cluster states of smaller size (shown so far at least for 1D cluster states)~\cite{dur03a}. These schemes must be investigated in more detail with respect to the proposed architecture before large-scale OWQC will be possible, but this system appears favorable for investigation of trapped-ion cluster states.

\section{Discussion \label{sec_discussion}}

The proposed method for quantum simulation of condensed-matter Hamiltonians may greatly reduce laser requirements, eliminating them for the interactions, but laser radiation will still be required for initialization and final-state measurement in most foreseeable ion QIP implementations.  Spontaneous emission is a beneficial requirement, not a hindrance, to laser cooling, optical pumping, and resonance fluorescence, all irreversible processes.  Though there are proposals for hot-ion quantum computing~\cite{james00a}, efficient state preparation using optical pumping and efficient state determination using resonance fluorescence are two of the biggest reasons trapped ions have been such a successful system for QIP.  They may not be easily removed from our simulation procedure, but their use can be relegated to the parts of the algorithm which do not require long-term coherence.  Additionally, individual addressing is not required for these steps, so the emission-dependent processes can be straightforwardly applied to the ion array using one beam and many mirrors, possibly on-chip micro-optics components~\cite{kimj05a,kimc07a,leibfried07a}, large laser sheets (beams whose waist in one direction is much larger than in the other) directed along the surface of the chip just above the surface, or even large cross-section beams encompassing the array and reflecting from the chip surface.

For the scheme's application to OWQC, where individual addressing for readout may be required, micro-optics can also be employed to switch the readout beam or beams to particular sites of the array, one at a time or in parallel.  Cooling and initialization can be done uniformly, as in the application to simulations.  Neither of these applications should require sympathetic recooling during calculation~\cite{rohde01a,blinov02a,barrett03a}, a probable necessity for large-scale circuit QC, since for the quantum simulations the ions are not moved until evolution is complete (if at all), and for OWQC, the ions are moved once, after which only the internal states are addressed (i.e., the possibly heated vibrational states are no longer relevant for the calculation).

Small ion-electrode distances are required for reasonable interaction strength in this scheme, and increased anomalous heating~\cite{turchette00a} could be problematic.  The 1--5~$\mu$m ion heights determined above to be required for strong interactions are at least a factor of 5 smaller than in traps in which heating has been examined to date, so the observed scaling behavior of heating with ion-electrode distance could limit interaction times.  There have been very promising results using cryogenically cooled traps~\cite{deslauriers06a,labaziewicz07a}, however, in which the anomalous heating rates have been shown to be reduced by orders of magnitude in a cooled trap structure.  This suggests that heating will not form an ultimate limit to interactions using this method if cooled electrodes are implemented.  Another area that must be explored for the successful implementation of this type of architecture is the vibrational mode structure of multiple ions in coupled microtraps.  If the strong-binding condition is satisfied, the phonons are generally localized to small regions around specific sites (i.e., they do not explore the entire array), and sideband cooling should thus not become untenable.  However, the degree to which microtrap frequencies must be equivalent, the general semilocalized vibrational mode structure in a microtrap array, and the method for efficient cooling are still to be determined.

The biggest advantage to using long-wavelength field excitation for the interactions is the almost complete elimination of error due to spontaneous emission and scattering.  However, there are other practical advantages to not requiring many optical beams.  The entire trap system may be miniaturized, and most of the important control fields can be brought in on wires or electromagnetic waveguides.  The interaction can be tailored to some degree through electrode configuration, a possibly more straightforward and reliable method than alteration of laser beam direction and intersection at the ions' location.  In addition, narrowband MW and RF signals can be produced at high powers much more easily than can optical radiation, maintaining relative phase coherence is greatly simplified, and intensity variation is much more controlled than in current high-power laser systems.  These considerations make a QC gate with error below the fault-tolerant level feasible using these methods. 

One disadvantage of the stimulated-Raman optical scheme that may not be eliminated with the architecture described here is the requirement of relatively high power (almost 1 W at a large detuning is required for fault-tolerant two-qubit gates using the Raman method~\cite{ozeri07a}).  The requirement for reasonably strong ion-ion interaction through RF magnetic-field-gradient excitation leads to relatively high input power to the microcoils in the trap structure.  For the interaction assumed above, a peak current of 10~mA is required in each (one-turn) microcoil.  At 2.5~mW per loop (50~$\Omega$ each), a 100$\times$100 array requires 25 W of RF power.  This large power requirement is not unrelated to that of the optical method.  It is somewhat less efficient in the RF case due to lack of focusing of the field from the surface-trap loops (as opposed to collimated laser beams) and the use of magnetic as opposed to electric dipole transitions, but the fact remains that the interaction Rabi frequency is proportional to field amplitude, and hence proportional to the square root of the power.  All large-scale QIP implementations, atomic or otherwise, will face this challenge when attempting to create strong interactions among large numbers of qubits, especially if large detunings are required to avoid scattering, as in the optical case.  A bright spot is that the power scaling in this architecture is linear with the number of spins or qubits in the system, so the exponential resource gain over classical simulations is not counterfeited.  In addition, the power to the coils can have a relatively low duty cycle, as it is required only during Hamiltonian evolution.  We note that, as part of the fabrication process a magnetically permeable material may be integrated at the center of each microcoil to create a stronger field above the chip, and depending on loss due to hysteresis and eddy currents, the power dissipation for a given interaction strength may be reduced.  

Despite the high power requirement, which is an engineering challenge that may be addressed through cryogenic operation or other methods, the architecture outlined here can allow significantly larger quantum simulations with much longer evolutions than would be the case in an optical scheme with easily attainable power.  The reduction of error from spontaneous scattering and emission could also make this architecture particularly attractive for one-way QC with trapped ions.  The optical properties of ions do not have to be a roadblock on the way to large-scale QIP.

\begin{acknowledgments}

We thank Malcolm Boshier and Jim Harrington for helpful discussions and comments on the manuscript.  This work has been supported
by the LANL Laboratory Directed Research and Development program.

\end{acknowledgments}

\bibliography{ion1}

\end{document}